\documentclass{article} 
\usepackage{jamaica04}
\frompage{000} \topage{000}                                              
\def\rightmark{$\boldmath{\mbox{$\phi$}}$ Meson Propagation}
\def\leftmark{K.~L. Haglin}
 
\title{$\boldmath{\mbox{$\phi$}}$ Meson Propagation and Decay at Finite 
Temperature}

\authors{
{Kevin L. Haglin 
}\\[2.812mm]
{\normalsize 
Department of Physics \& Astronomy, Saint
Cloud State University \\ 
720 Fourth Avenue South, St. Cloud, MN 56301  USA\\[0.2ex] 
}}
 
\abstract{We study spectral properties of the $\phi$ meson
at finite temperature using an effective Lagrangian together
with finite-temperature field theory as a basis for modeling. 
General field-theoretic arguments are then used to establish
the lifetime of $\phi$ in hot hadronic matter.  We find
from the model, and we therefore propose a scenario in which,
the phi decays inside the fireball.  Early decays into $\mu^{+}\mu^{-}$
occur at high temperature while measurable
hadronic decays into {\it\,K\/}$^{+}${\it\,K}\/$^{-}$ (not
suffering rescattering) occur at freezeout, where
flow could be substantial.  These results provide a consistent
picture, if not possible interpretation of NA49 and NA50 data from CERN.
         }
 
\begin{document}
 
\maketitle
\setcounter{page}{1}

\section{Introduction}\label{intro}
Dynamics of strongly-interacting many-body systems created
in high-energy heavy-ion collisions are nontrivial owing to
such features as the time and distance scales involved, the unavoidable 
nonperturbative component, and the quantum mechanical
implications of multiparticle coherence.  At present, the
challenges go beyond the reach of a complete theory.  Instead, we
focus on specific particle signals to probe the overall dynamics.  Probes 
are carefully selected as each one provides different opportunities to extract
information and allow bridging between an observable final 
state and the unobservable collision history.
Electromagnetic signals are well suited for studying the
early stages since they do not suffer reinteractions
once produced, while hadronic messengers typically bring
forward information on the freezeout stage since they
rescatter.
Vector mesons are especially good
tools for probing the systems.  Indeed, the phi meson is
both desirable and somewhat undesirable since its
vacuum lifetime is $\sim$ 45 fm/{\it\/c\/}, which is longer
than the expected fireball duration.   It is desirable
since its mass is intermediate between $\rho$ and
{\it\,J\/}/$\psi$ and can thus be identified cleanly. 
The relatively long lifetime, however, means
that it is less able to probe the early stages of the
system where hadronic physics meets subhadronic (quark) degrees of
freedom. 

Still, the phi is pursued since it decays both into hadronic
final states (kaon pairs) and electromagnetic states (muon
pairs).  Even though the vacuum lifetime is long, there is
hope that some $\phi$'s will decay inside and fingerprint the 
hot medium and its collective nuclear properties.  Recent
experimental results from CERN have reported charged kaon
pair signals and muon pair signals in two separate experiments
(NA49 and NA50)\cite{na49,na50}.   The results have been 
contemplated and have been
dubbed ``the $\phi$ puzzle'', because the kaon spectra
seem to show an inverse slope parameter greater than the
muon spectra.  This could mean that the kaons
are showing effects of strong flow, while the muons
are perhaps not.  But this seems inconsistent.  If the
phi meson decays outside the fireball (long lifetime for phi),
then either both signals should sit on top of the strong flow
foundation, or both should not.  
So the $\phi$ puzzle presents a serious theoretical challenge.

Our article is organized in the following way.  We discuss in 
Sect.~\ref{chirall} the details of the effective
Lagrangian and the formalism from finite-temperature field
theory used to compute the spectral function. 
Sect.~\ref{drate} includes a brief discussion of the decay rate 
at finite temperature. Results for $\phi$ lifetime are presented and 
discussed in Sect.~\ref{results} 
with a proposed scenario to interpret experimental results. 
Finally, Sect.~\ref{concl} summarizes and concludes.

\section{Effective Lagrangian and Spectral Function}\label{chirall}  
Owing to the strangeness content of the $\phi$, the dominant
decay channel is through kaon pairs.  The interaction Lagrangian
we use to describe this is
\begin{eqnarray}
{\cal\,L\/}_{\phi\/K\/K} & = & g_{\phi\,K\/K}\left(\partial_{\mu}{\bf\/K}\times
{\bf\/K\/}\right)_{(3)}\,\phi^{\mu}\/,
\quad
{\cal\,L\/}_{\phi\/\phi\/K\/K} \ = \  
g_{\phi\,K\/K}^{2}\,\left|\phi^\mu\,{\bf\/K}\,\right|^{2}\/,
\label{ell1}
\end{eqnarray}
where the subscript (3) indicates that isospin picks off the
neutral component of the cross product.
Eq.(\ref{ell1}) represents a mere subset of the
interactions generated from a complete three-flavor
chiral Lagrangian.  For now, this is all which is needed.
However, there are OZI suppressed decays into $\pi\rho$ which
we model with the following
\begin{eqnarray}
{\cal\,L\/}_{\phi\pi\rho} & = & g_{\phi\pi\rho}\,\epsilon_{\mu\nu\alpha\beta}\,
\partial^{\mu}\phi^{\nu}\partial^{\alpha}{\boldmath{\mbox{$\rho$}}}^{\/\/\beta}
\boldmath{\mbox{$\cdot$}}\,{\boldmath{\mbox{$\pi$}}}.
\label{ell3}
\end{eqnarray}
Coupling constants are fitted to 
observed on-shell
decays
$\phi\rightarrow\/K^{+}\/K^{-}, \phi\rightarrow\pi\/\rho$\cite{pdg}.

The interactions identified in Eqs.~(\ref{ell1}) and (\ref{ell3})
give rise to self-energy contributions from kaon bubble and tadpole graphs
and from a $\pi$--$\rho$ loop diagram.  In Euclidean metric then,
\begin{eqnarray}
\Pi^{\mu\nu}\left(k\/\right) & = &
g_{\phi\/KK\/}^{2}\,T\/\sum\limits_{p_{4}}\int\,{d^{3}\/p\over
\left(2\pi\/\right)^{3}}
\left\lbrack
\,{-(2p+k)^{\mu}(2p+k)^{\nu}\over(p^{2}+m_{K}^{2})\left[
(p+k)^{2}+m_{K}^{2}\right]}
+ 
\,{2\/\delta^{\mu\nu}\over(p^{2}+m_{K}^{2})}
\right\rbrack
\nonumber\\
& \ & 
+
g_{\phi\pi\rho}^{2}\,T\/\sum\limits_{p_{4}}\int\,{d^{3}\/p\over
\left(2\pi\/\right)^{3}}
\,{N^{\mu\nu}\over(p^{2}+m_{\pi}^{2})\left[
(p+k)^{2}+m_{\rho}^{2}\right]}\/,
\end{eqnarray}
where 
\begin{eqnarray}
N^{\mu\nu} & = & 
\left[\left(k\cdot\/p\right)^{2}-p^{2}k^{2}\right]\,\delta^{\mu\nu}
+p^{2}\,k^{\mu}k^{\nu}
+k^{2}\,p^{\mu}p^{\nu}
-\left(k\cdot\/p\right)\left[p^{\mu}k^{\nu}+
k^{\mu}p^{\nu}\right].
\end{eqnarray}
The {\it\/T\/} = 0 expressions are handled with dimensional regularization
\cite{ramond}, while at {\it\,T\/} $>$ 0
the discrete sum over Matsubara frequencies ($p_{4} = 2\pi\,T\times$ integer)
is conveniently
carried out using contour integration.  Details will not be reproduced
here, but do appear already in the literature\cite{cgjk91,khcg94}.  We
point out that the self-energy has real and imaginary parts 
which are both temperature and momentum dependent. Further, polarization 
effects have been shown to be small.

Two-loop contributions are numerous.  The imaginary parts
of the graphs correspond physically to scattering processes.
Leading contributions are these: $\phi+K\rightarrow\phi+K$,
$\phi+\pi\rightarrow\,K^{*}\/K$,
$\phi+K^{*}\rightarrow\,\pi\/K$, etc.  The imaginary
part of the self energy at two loops is most
easily computed within kinetic theory recognizing generally that\cite{aw83}
\begin{eqnarray}
\mbox{Im}\Pi & = & -\omega\,\Gamma^{\rm\/coll}.
\end{eqnarray}
The differential collision rate for the process
$\phi + b \rightarrow 1 + 2$ is 
\begin{eqnarray}
d\Gamma^{{\rm\/coll}} & = &
{g\over n_{\phi}} 
{d^{3}p_{\phi}\over(2\pi)^{3}2E_{\phi}}f_{\phi}
{d^{3}p_{b}\over(2\pi)^{3}2E_{b}}f_{b}
{d^{3}p_{1}\over(2\pi)^{3}2E_{1}}(1+{f}_{1})
{d^{3}p_{2}\over(2\pi)^{3}2E_{2}}(1+{f}_{2})
\nonumber\\
 & \ & \times |\bar{\cal\/M\,\,}\hskip-0.2em|^{2}\,(2\pi)^{4}\,\delta^{4} 
\left(p_{\phi}+p_{b}-p_{1}-p_{2}\right)\/.
\end{eqnarray}
Higher-order effects are handled by introducing
monopole form factors at all vertices.
A long list of contributions has been studied
for mesons\cite{kh95,volker02}
and baryons\cite{wskh98}.   The net
result for the collision rate is an increasing function of 
temperature; the increase is especially rapid past 
150 MeV and we note that {\it\/K\/}$^{*}$(892) plays
an important role\cite{volker02}.

\begin{figure}[htb]
\vspace*{-0.15cm}
                 \insertplot{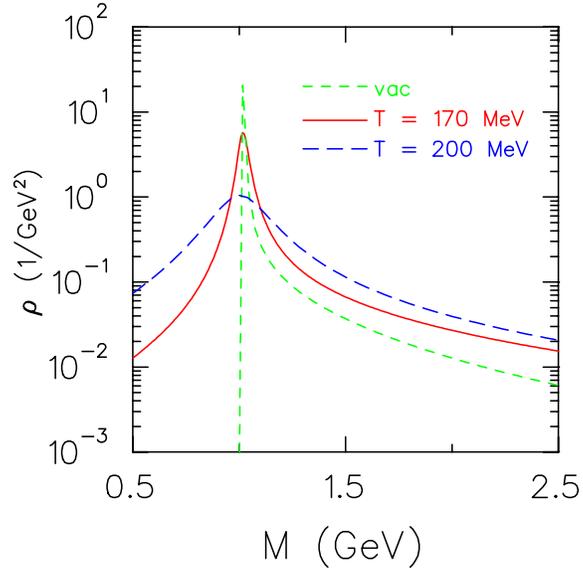}
\vspace*{-0.5cm}
\caption[]{Spectral function for $\phi$ meson in the vacuum
(short-dashed), at {\it\/T} = 170 MeV (solid) and at {\it\/T} = 200 MeV
(long-dashed) curves.}
\label{fig1}
\end{figure}

We write the spectral function as 
\begin{eqnarray}
\rho(M) &= & {\mbox{1}\over\pi}\,{\mbox{--\,Im}\Pi\over
(M-m_{\phi}^{2}-\mbox{Re}\Pi\/)^{2}+(\mbox{Im}\Pi\/)^{2}}\,.
\end{eqnarray}
At low temperatures the imaginary part of the self energy
is dominated by decays.  At temperatures of the order of
the pion mass and higher, the imaginary part is dominated
by scattering in the hadronic medium.  We plot in Fig.~\ref{fig1} 
the vacuum spectral function as well as results
for two relevant temperatures.  Notice how strongly distorted the
spectral properties become as the temperature reaches
a value where deconfinement might be expected.

\section{$\boldmath{\mbox{$\phi$}}$ Decay at $\boldmath{\mbox{$\,T\/$}}$ 
$\boldmath{\mbox{$>$}}$ 0}\label{drate}  

Quite generally, a resonant hadronic state $\left|R\,\right\rangle$
decays into the multiparticle final state 
$\left|f\,\right\rangle$ = $p_{1}p_{2}\ldots\,p_{\ell}$ at the following
rate\cite{aw83}
\begin{eqnarray}
{d\/N_{f}\over\/d^{4}x\,d^{4}q\/}
& = & 
{(2J+1)\over(2\pi)^{\/3}}{1\over\exp(\beta\/q_{0})\pm\/1}
\left\lbrack
\rho(M)
\right\rbrack\,2M\,\Gamma^{{\rm\/vac}}_{R\rightarrow\,f\/}.
\label{fourrate}
\end{eqnarray}
We integrate Eq.~(\ref{fourrate}) over all three momentum and
all off-shell energies to obtain the number of
decays per unit time per unit volume.
From there, we simply
divide by the number of $\phi$'s per unit volume to arrive
at the number of decays per unit time. 
Finally then, the finite temperature lifetime is the
inverse of the rate.
\begin{eqnarray}
\tau_{\phi} & = & 
{1\over\Gamma^{\rm\,decay}} =  \left\lbrack{1\over\,n_{\phi}}\,
{d\/N_{f}\over\/d^{4}x\/}\right\rbrack^{\/-1}.
\label{taudefined}
\end{eqnarray}

\section{Results}\label{results}  

\subsection{Lifetime}\label{tau}

\begin{figure}[htb]
\vspace*{0.1cm}
                 \insertplot{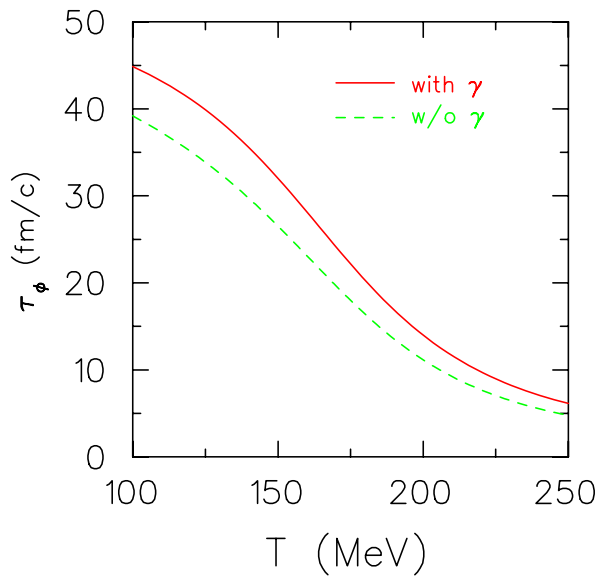}
\vspace*{-0.5cm}
\caption[]{Lifetime for $\phi$ at finite temperature. Solid curve
is Eq.~(\ref{taudefined}), dashed curve results from dividing out a
thermal Lorentz factor.}
\label{fig2}
\end{figure}

The lifetime for $\phi$ at finite temperature is
plotted in Fig.~\ref{fig2}.  The first notable feature is
that it is a monotonically decreasing function of temperature.
This is not trivial, since the decay rate has a decreasing tendency
from a Lorentz time-dilation factor.  But the dominant 
effect is the broadening of the spectral function which opens up
phase space faster than effects which tend to
decrease the rate.  In the evolution of the fireball, at temperatures
where it begins to make sense to use hadronic language
to discuss dynamics, the lifetime has dropped in the model
to 5--8 fm/{\it\/c\/}.  This is noteworthy.  It means
that at high temperatures, the $\phi$ seems to decay inside the fireball!

The picture which begins to emerge is that many $\phi$'s decay
inside the fireball---both to kaon pairs and to muon pairs.
The muon pairs escape the system and travel to the detectors.
The kaons, on the other hand, do not escape since at that temperature
the mean free path is on the order of 1 fm\cite{khsp94,volker02}.  The 
hadronic system expands and cools until the mean free path is of the 
order of the size of the system.  Studies have shown that a reasonable
temperature for this is around the pion mass or below.  By that
stage, flow has had a good chance to build up.  We therefore
explore in the next section the expected kaon-pair momentum
spectrum at freezeout, but in the face of strong flow.

\subsection{Transverse mass distribution}\label{d2mtdmtdy}
 
\begin{figure}[htb]
\vspace*{0.1cm}
                 \insertplot{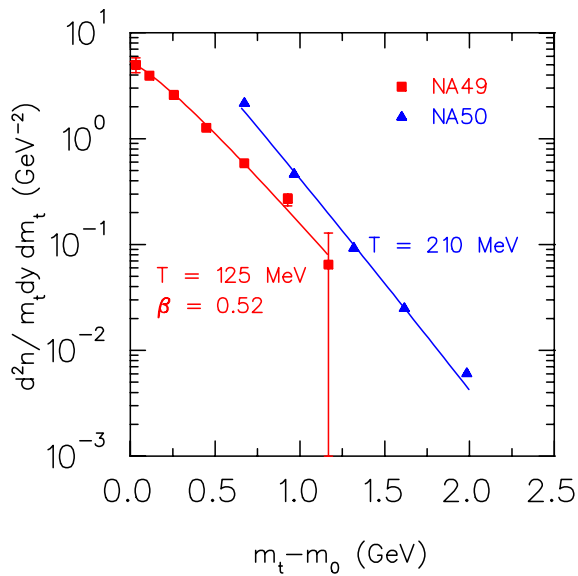}
\vspace*{-0.5cm}
\caption[]{Differential {\it\/m\/}$_{t}$ distribution from Eq.~(\ref{sr})
at {\it\/T} = 210 MeV with zero flow, and {\it\/T} = 125 MeV 
with {\it\/v} = 0.52{\it\/c} compared
with experimental results from CERN.  Particle yields in the model are not
fixed.}
\label{fig3}
\end{figure}

In this section we take the Siemens-Rasmussen formula
for radial flow\cite{sr79} and generalize to include
quantum statistics.  In the local rest frame we take an
equilibrium distribution (Bose or Fermi) and assume
the matter is moving radially outward with velocity {\it\/v\/}. Next,
we Lorentz transform back to the fireball rest frame, and finally,
integrate out the unobserved variables (and evaluate at 
zero rapidity).  Our result is
\begin{eqnarray}
{d^{2}n \over m_{t}\,dm_{t}\,dy\/} & = & 
\sum\limits_{\ell=1}^{\infty}
{(\mp)^{(\ell+1)}
\over(2\pi)^{2}}
\,e^{-\gamma\/m_{t}\ell/T}
\left\lbrack
\left(\gamma\/m_{t}+{T\over\ell}\right)\/{\mbox{sinh}(\alpha\ell)
\over{\alpha\ell}}
-{T\/\mbox{cosh}(\alpha\ell)\over\ell}
\right\rbrack,\quad\quad
\label{sr}
\end{eqnarray}
where the + (--) sign corresponds to bosons (fermions).
Of course, the first term (the Boltzmann limit) dominates unless
flow becomes very high or masses are very low.  Neither circumstance
occurs here, so the Boltzmann limit is sufficient.  We plot in
Fig.~\ref{fig3} the transverse mass spectra using reasonable values
for freezeout temperature and flow as compared with the
experimental results from the NA49 and NA50 collaborations. 

\section{Conclusions}\label{concl}
We have modeled interactions between the $\phi$ and light
mesons using a three-flavor chiral Lagrangian.  The one- and
two-loop contributions to the $\phi$ self energy were studied
with respect to their influence on the spectral function
at finite temperature.  The $\phi$ spectral properties were
observed to be significantly modified as the temperature reached 
the expected crossover point to deconfined quark matter.
However, a modified
spectral function does not immediately and directly imply
anything particular about the {\it lifetime\/} of the state in
hot hadronic matter.  Lifetime estimation requires separate analyses.

We have therefore used the finite temperature spectral function
as input to a generalized Breit-Wigner distribution,
folded with free-space decay dynamics, in order to estimate the
decay rate at finite temperature.  Trivially, the lifetime
is its inverse.  We find that the lifetime of
the phi meson at high temperature is roughly 5--8 fm/{\it\/c\/}.
This is short enough where decay inside the fireball is
quite likely. 

Finally then, we propose a scenario in which the
phi decays early in the heavy-ion collision---at high temperature.  Only
the electromagnetic channel survives from here to the final state.
Kaon rescattering is too aggressive for them to escape.   Later however, 
when the temperature has dropped to a value, say, below the pion mass,
and flow has had a chance to build up strongly, the kaons from
phi decay will reach the detector.  Of course at that stage, the free-space
mass and width for $\phi$ should dominate.  The Siemens-Rasmussen
formula for radial flow was generalized to include
effects beyond the usual Boltzmann distribution.  The values we find for 
temperature and flow when comparing to the NA49 and NA50
data are the following.  Muon pair production is dominated by temperatures
around 210 MeV, while the kaon distribution seems to be consistent
with a temperature of 125 MeV and a flow velocity of {\it\/v} = 0.52{\it\/c}.
These values do not reflect an exhaustive chi-square search through
parameter space, but are only indications of what is possible.  A
more complete analysis, including branching ratio studies, will be published 
elsewhere\cite{khlh}.

\section*{Acknowledgement}
This work was supported in part by the National Science
Foundation under grant number PHY-0098760.
 

\vfill\eject
\end{document}